\documentclass[graybox]{svmult}

\usepackage{mathptmx}       
\usepackage{helvet}         
\usepackage{courier}        
\usepackage{type1cm}        
\usepackage{makeidx}         
\usepackage{graphicx}        
\usepackage{multicol}        
\usepackage[bottom]{footmisc}

\makeindex             

\def\mincir{\raise -2.truept\hbox{\rlap{\hbox{$\sim$}}\raise5.truept \hbox{$<$}\ }}
\def\mincireq{\hbox{\raise0.5ex\hbox{$<\lower1.06ex\hbox{$\kern-1.07em{\sim}$}$}}}
\def\magcir{\raise-2.truept\hbox{\rlap{\hbox{$\sim$}}\raise5.truept \hbox{$>$}\ }}

\begin{document}

\title*{Nonthermal Emission from Star-Forming Galaxies}
\author{Yoel Rephaeli and Massimo Persic}
 
\institute{Yoel Rephaeli \at School of Physics and Astronomy, Tel Aviv University, 
Tel Aviv, 69978, Israel\\and\\
Center for Astrophysics and Space Sciences, University 
of California, San Diego, La Jolla, CA 92093-0424, USA, \email{yoelr@wise.tau.ac.il}
\and Massimo Persic \at INAF/Osservatorio Astronomico di Trieste and INFN-Trieste, 
via G.B.Tiepolo 11, I-34143 Trieste, Italy \email{persic@oats.inaf.it}}
%
%
%
\maketitle

\abstract{
The detections of high-energy $\gamma$-ray emission from the nearby starburst 
galaxies M\,82 \& NGC\,253, and other local group galaxies, broaden our knowledge 
of star-driven nonthermal processes and phenomena in non-AGN star-forming galaxies. 
We review basic aspects of the related processes and their modeling in starburst 
galaxies. Since these processes involve both energetic electrons and protons 
accelerated by SN shocks, their respective radiative yields can be used to explore 
the SN-particle-radiation connection. Specifically, the relation between SN 
activity, energetic particles, and their radiative yields, is assessed through 
respective measures of the particle energy density in several star-forming 
galaxies. The deduced energy densities range from ${\cal O}(10^{-1})$\,eV\,cm$^{-3}$ 
in very quiet environments to ${\cal O}(10^2)$\,eV\,cm$^{-3}$ in regions with very 
high star-formation rates.}

\section{Introduction}
\label{sec:1}

%

High star formation (SF) and supernova (SN) rates in starburst (SB) 
galaxies (SBGs) boost the density of energetic nonthermal particles, 
whose main constituents are protons and electrons. Coulomb, synchrotron 
and Compton energy losses by the electrons, and the decay of pions following 
their production in energetic proton interactions with protons in the ambient 
gas, result in emission over the full electromagnetic spectrum, from radio to 
very high-energy (VHE, $\geq$100\,GeV) $\gamma$-rays. The relatively high 
intensity emission in SBGs, as compared with emission form `normal' 
star-forming galaxies (SFGs), makes nearby members of this class the most 
likely non-AGN targests for $\gamma$-ray telescopes, such as {\it Fermi} 
and the (Cherenkov arrays) H.E.S.S., MAGIC, and VERITAS.

Interest in $\gamma$-ray emission from SFGs clearly stems from the prospects 
for improved understanding of the origin and propagation mode of energetic 
electrons and protons and their coupling to interstellar media. This interest 
has been enhanced by recent detections of the two nearby SBGs M82 \& NGC253 by 
{\it Fermi} (Abdo et al. 2010a) and, respectively, by H.E.S.S ((Acciari et al. 
2009) and VERITAS (Acero et al. 2009). M31, the closest normal spiral 
galaxy, was also detected by {\it Fermi} (Abdo et al. 2010b). 

A realistic estimate of the expected $\gamma$-ray emission requires a detailed 
account of all relevant energy loss processes of energetic electrons and protons 
as they move out from the central SB source region into the outer galactic disk. 
Calculations of the predicted X-$\gamma$-ray spectra of nearby galaxies were made 
long ago with varying degree of detail (e.g., Goldshmidt \& Rephaeli 1995, Paglione 
et al. 1996, Romero \& Torres 2003, Domingo-Santamar\'\i a \& Torres 2005). A more 
quantitative numerical approach was initiated by Arieli \& Rephaeli (2007, 
unpublished), who used a modified version of the GALPROP code (Moskalenko \& 
Strong 1998, Moskalenko et al. 2003) to solve the Fokker-Planck diffusion-convection 
equation (e.g., Lerche \& Schlickeiser 1982) in 3D with given source distribution and 
boundary conditions for electrons and protons. This numerical treatment was 
implemented to predict the high-energy spectra of the two nearby galaxies M82 
(Persic, Rephaeli, \& Arieli 2008, hereafter PRA) and NGC253 (Rephaeli, Arieli, 
\& Persic 2010, hereafter RAP). The predictions made in these papers agree well 
with observations made with Fermi and TeV arrays, as will be discussed in the 
next section. 

Particle acceleration and propagation in galactic environments are largely similar 
in all SFGs. What mainly distinguishes a SBG from a normal SFG is the dominance of 
a relatively small central region of intense star formation activity. The overall 
validity of the numerical treatments of the two nearby SBGs provides a solid basis 
for generalizing the model to SFGs in general. 

We briefly review the calculation of steady-state particle spectra and their 
predicted radiative spectra for the above two nearby SBGs, and discuss similar 
calculations for conditions in a SFG. The particle energy density can be 
determined in several different ways. In order to assess and gauge the SN-energetic 
particle connection we compare estimates of the energetic proton (which dominate 
the) energy density in SFGs by three different methods, finding overall agreement, 
which provides further evidence for the validity of the basic approach.

\section{Particle and Radiation Spetra in Starburst Galaxies}
\label{sec:2}

Acceleration in SN shocks by the first-order Fermi process yields a power-law 
distribution with index $q$$\geq$2 (e.g., Protheroe \& Clay 2004) in a very wide 
energy range, from a value close to the mean thermal energy of the gas particles 
(in non-relativistic shocks) to a very high value ($\geq$10$^{14}$ eV). The 
accelerated proton-to-electron (p/e) density ratio, $N_{\rm p}/N_{\rm e}$, in 
the source (either the SB or the full disk) region can be calculated assuming 
charge neutrality (Bell 1978, Schlickeiser 2002). This ratio reaches its maximum 
value, $(m_{\rm p}/m_{\rm e})^{(q-1)/2}$ (for $q$$>$1; $m_{\rm e}$ and $m_{\rm p}$ 
are the electron and proton masses), over most of the relevant range of particle 
energies, $E$$>$1\,GeV. (For the dependence of this ratio on particle energy, 
and more discussion on this and other relevant physical processes, see PRA and 
references therein.) 

The electron density in the source region, $N_{\rm e}$, is inferred from radio 
measurements (of the same region); by adopting the theoretically expected 
expression $N_{\rm p}/N_{\rm e}=(m_{\rm p}/m_{\rm e})^{(q-1)/2}$, the proton 
density $N_{\rm p}$ can be deduced. The fit to the radio data provides both the 
normalization of the electron spectrum and the {\it actual} value of $q$, which 
is found to be somewhat larger than $2$, even in the central SB region. In this 
procedure the electron population is composed of both primary and secondary 
electrons, with the latter self-consistently determined by accounting for the 
pion yield of energetic protons with protons in the gas. We note that the 
theoretically predicted value of the density ratio is valid in the source region, 
where energy equipartition is more likely to be attained since the relevant 
processes couple particles and fields more effectively than in the rest of galactic 
disk.

The particle spectral distributions evolve differently as they propagate 
out from their acceleration region. Typically, the electron spectrum is  
most directly deduced from  measurements of synchrotron radio emission. 
The inferred spectrum can be related to the source spectrum through a solution 
of the kinetic equation describing the propagation modes and energy losses by 
electrons and protons as they move out from their acceleration region. 
A very useful detailed description of the time-dependent spectro-spatial 
distribtion of protons, diffusing out of a region with a discrete population 
of acceleration sites, was recently given by Torres et al. (2012). This 
study elucidates the explicit dependence of the distribution on distance 
from the acceleration site, energy loss time, and the diffusion coefficient. 
It also follows the temporal evolution of the distribution towards a steady state.

Since the estimated duration of a SB phase is $\sim 10^8$ yr, a timescale 
which is much longer than any of the relevant energy loss or propagation 
timescales for electrons and protons, in (essentially) all treatments a 
steady state is assumed to be attained. Since the calculation of particle 
steady-state spectra requires inclusion of all the important energy loss 
mechanisms and modes of propagation, the treatment is necessarily numerical. 
We have employed the code of Arieli \& Rephaeli (2007), which is based on a 
modified version of the GALPROP code (Moskalenko \& Strong 1998, Moskalenko 
et al. 2003), to solve the kinetic equation for $N_{i}(\gamma, R,z)$, where 
$i=e, p$, $\gamma$ is the Lorentz factor, and $R$ and $z$ are the 2D spatial 
radius and the coordinate perpendicular to the galactic plane. The exact 
Fokker-Planck diffusion-convection equation (e.g., Lerche \& Schlickeiser 
1982) was solved in 3D with given source distribution and boundary conditions 
for electrons and protons. In addition to diffusion with an energy dependent 
coefficient, particles are assumed to be convected by a galactic wind with 
spatially varying velocity.

The dominant energy losses of high-energy electrons are synchrotron emission 
and Compton scattering by the FIR and optical radiation fields; these processes 
(obviously) depend on the mean strength of the magnetic field, $B$, and the 
energy density of the radiation fields, respectively. At energies below few 
hundred MeV, electrons lose energy mostly by Coulomb interactions with gas 
particles. At low energies proton losses are dominated by Coulomb 
interactions with gas particles. Protons with kinetic energy above the 
(range of) pion masses ($\sim$140 MeV) lose energy mainly through interactions 
with ambient protons, yielding neutral ($\pi^{0}$) and charged ($\pi^{\pm}$) 
pions. Neutral pions decay into photons, while decays of $\pi^{\pm}$ result 
in energetic e$^{\pm}$ and neutrinos. The proton and (total) electron components 
are coupled through the production of secondary electrons in $\pi^{-}$ decay 
(following their creation in {\it pp} interactions).

Measured synchrotron radio spectra provide the critically important information 
on the particle spectra and their overall normalization: Fitting the predicted 
radio emission to measurements fixes normalization of the steady state electron 
and - based on a theoretical prediction - proton energy distributions. From 
these measurements alone the electron density and mean magnetic field cannot 
be separately determined. To do so it is usually assumed that particle and 
magnetic field energy densities are equipartitioned. In our numerical 
treatment this approach necessitates an iterative procedure to solve for 
$N_{\rm e}$, $N_{\rm p}$, and the field strength at the center, $B_0$, given 
a measured value of the radio flux. 

Particles diffuse and are convected out of their source region. Diffusion is likely 
to be random walk against magnetic field inhomogeneities, with an estimated value 
of $\sim 3 \times 10^{28}$ cm$^2$/s for the central diffusion coefficient. Convection 
is by galacitc wind with a typical velocity of $\sim 500$ km $s^{-1}$ (Strickland et 
al. 1997) in the source region. Based on Galactic cosmic-ray MHD wind models, we 
assume that the convection velocity increases linearly with distance from the disk 
plane (e.g. Zirakashvili et al. 1996). 

The other quantities needed to calculate the steady state distributions 
of electrons and protons are the densities of neutral and ionized gas in the 
central SB region and throughout the disk, the central value of the (mean) 
magnetic field and its spatial profile across the disk, and energy densities 
of ambient radiation fields (including the CMB). As discussed in PRA, it 
is assumed that magnetic flux is conserved in the IS ionized gas, so that the mean 
strength of the field can be related to the local ionized gas density, 
$n_{\rm e}$, using the scaling $B \propto n_{\rm e}^{2/3}$ (Rephaeli 1988). 
If instead energy equipartition is assumed, and the magnetic energy density is 
scaled to the thermal gas energy density, then the proportionality relation is 
$B \propto n_{\rm e}^{1/2}$. In our work we have taken the ionized gas density 
profile to be $n_{e} \propto exp(-z/z_0)/(1 + (R/R_0)^2)$, typically with $R_0=1.5$ 
kpc, and $z_0 =0.5$ kpc (as deduced for NGC\,253 by Strickland et al. 2002). 

\begin{figure}[h]
\label{fig:radio_spectrum}
\centering
\includegraphics[scale=.45]{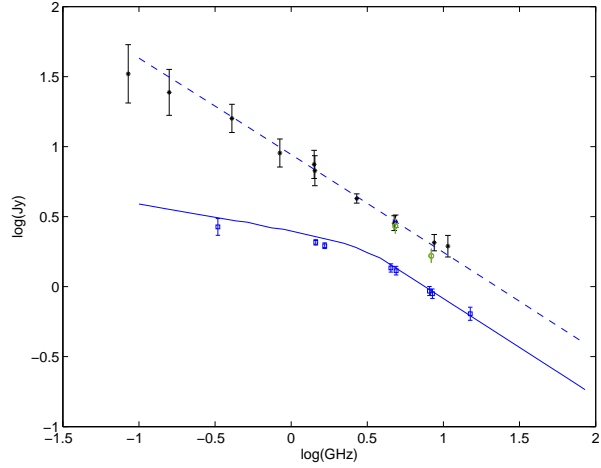}
\caption{Spectral fits to radio measurements of the SB and entire disk 
regions of NGC\,253 (RAP). The solid line is a fit to the emission from the SB 
region; the dashed line is a fit to the emission from the entire disk. Data 
are from Klein et al. (1983, black dots), Carilli (1996, blue squares), and 
Heesen et al. (2008, green circles).}
\end{figure}

Given the measured radio fluxes from the central and full disk regions of 
the two nearby SBGs M82 and NGC253, shown in Figure 1 for the latter galaxy, 
and values of all the above quantities, the steady state particle spectra 
and their radiative yields were calculated using the modified GALPROP code. 
Here we present the results of this work; more details on the method and 
values of the input parameters can be found in PRA and RAP.

\begin{figure}[h]
\label{fig:ep_density}
\centering
\includegraphics[scale=.40]{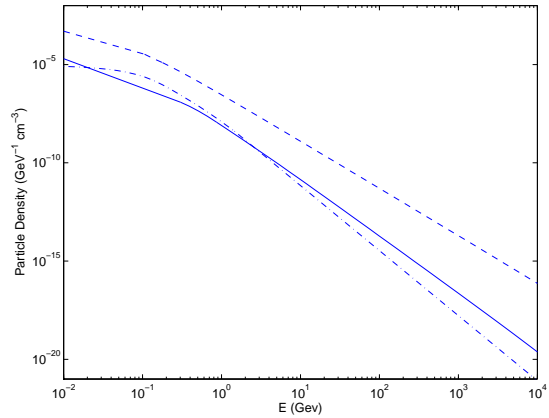}
\caption{Primary proton (dashed line), primary electron (solid line), and 
secondary electron (dashed-dotted line) spectral steady state density 
distributions in the central SB region NGC\,253 (RAP).}
\end{figure}

The high-energy photon spectra of NGC253 are shown in Figure 3 (from RAP). 
Emission levels depend mostly on values of the proton to electron ratio in the 
source region, on the magnetic field, gas density, and their spatial profiles. The 
basic normalization of the electron density is provided by the measured 
radio emission in the source region; the variation of the electron 
spectrum across the disk is largely determined by synchrotron losses. 
Uncertainty in the estimated level of emission is largely due to the steep 
dependence of the electron density on the field. As argued by RAP, the central 
value of the (mean) magnetic field is unlikely to be appreciably higher than 
the value deduced in N253 (and also in M82), $B_{0} \sim 200 \, \mu$G. A lower 
field value would result in a reduced proton density and a lower rate of $\pi^0$ 
decays. For a given radio flux the electron density would have to be correspondingly 
higher, resulting in higher bremsstrahlung and Compton yields, roughly compensating for 
the lower level of hadronic emission. Emission from $\pi^0$ decay depends 
linearly on the ambient proton density in the central disk region.

\begin{figure}[h]
\label{fig:spectra_all}
\centering
\includegraphics[scale=.40]{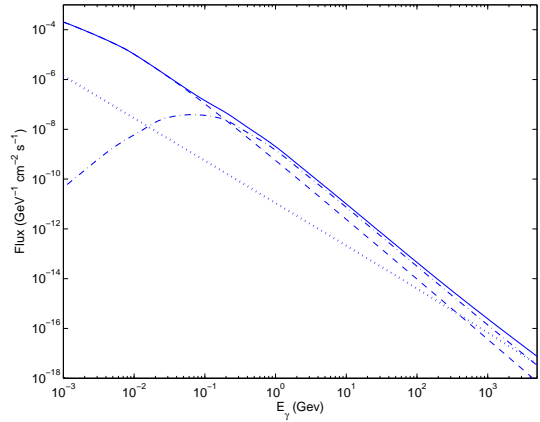}
\caption{High-energy emission from the disk of NGC\,253. 
Radiative yields are from Compton scattering off the FIR radiation field 
(dotted line), electron bremsstrahlung off ambient protons (dashed line), 
$\pi^0$ decay (dashed-dotted line), and their sum (solid line).}
\end{figure}

\begin{figure}[h]
\label{fig:spectra_all}
\centering
\includegraphics[scale=.40]{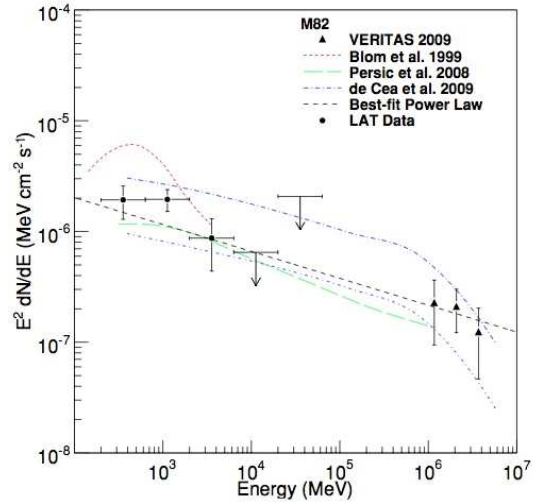}
\caption{High-energy emission from M82. VERITAS (Acciari et al. 2009) 
and {\it Fermi} measurements are shown together with predicted spectra. 
Figure is reproduced from Abdo et al. (2010a).}
\end{figure}

\begin{figure}[h]
\label{fig:spectra_all}
\centering
\includegraphics[scale=.42]{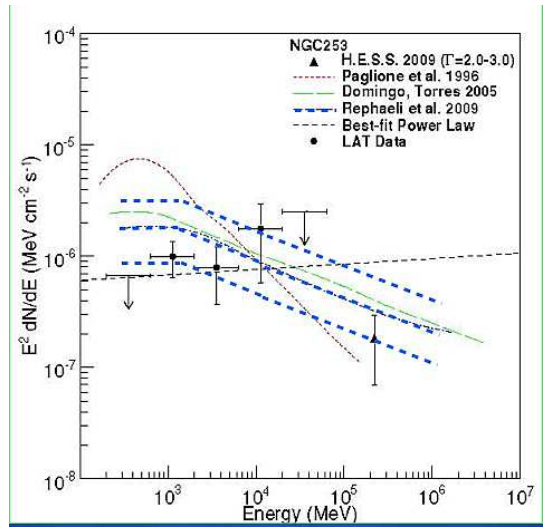}
\caption{High-energy emission from the disk of NGC\,253. {\it Fermi} and 
H.E.S.S measurements are shown together with predicted spectra. The figure 
is reproduced from Abdo et al. (2010b); the added dashed blue lines show our 
predicted spectrum with the estimated range of $1\sigma$ uncertainty.}
\end{figure}

Detailed modeling of high-energy emission from M82 by PRA preceded its 
detection by 
VERITAS (Acciari et al. 2009) and {\it Fermi} (Abdo et al. 2010a). 
These observations are shown in Figure 4 (from Abdo et al. 2010a) , 
together with theoretical predictions including that of PRA. The detected 
flux level agrees well with that predicted by PRA and by 
de Cea et al. 2009. Results from our similar treatment of the steady-state 
particle and radiation spectra of NGC253 were presented in RAP. The estimated 
high energy fluxes for this galaxy (RPA, Paglione et al. 1996, Domingo-Santamaria 
\& Torres 2005) were also in the range measured by {\it Fermi} (Abdo et al. 
2010a), and - in the TeV region - by the HESS telescope (Acero et al. 2009), given 
the substantial observational and modeling uncertainties. 
These observational results and predicted spectra are shown in Figure 5 
(adopted from Abdo et al. 2010a); our integrated spectrum and estimated 
$1\sigma$ uncertainty region are marked by the dashed blue lines (inserted 
into the original figure). 
These two SBGs are the only extragalactic non-AGN sources that were detected at both 
GeV and TeV energies. The FIR luminosities of these galaxies are at or somewhat below 
the nominal level for SBGs; clearly, their apparent high brightness is due to their 
proximity. 

A few other local and nearby galaxies were also detected at high energies: LMC 
(Abdo et al. 2010c), SMC (Abdo et al. 2010d), Andromeda (M31; Abdo et al. 2010b), 
and the composite Sy2/SB galaxies NGC\,1068 and NGC\,4945 (Lenain et al. 2010). 
The emission in these galaxies is mostly hadronic $\gamma$-ray emission, except 
for NGC\,1068 where emission from the active nucleus may be dominant. Among 
galaxies whose detected high energy emission is powered by stellar activity, 
Arp\,220 has the highest SFR in the local universe, yet an attempt to detect 
it by MAGIC (Albert et al. 2007) was not successful due to its relatively large 
distance.

\section{Estimates of Cosmic-Ray Energy Density}
\label{sec:3}

Active star formation in galaxies leads to acceleration of protons and electrons 
via the Fermi-I diffusive shock acceleration mechanism in SN remnants. 
Under equilibrium conditions in a galaxy, a minimum-energy configuration of the 
magnetic field and the energetic particles may be attained. Energy densities of 
particles and magnetic fields may then be in approximate equipartition, implying 
that the energetic proton energy density, $U_{\rm p}$, can be deduced from the 
detected level of synchrotron radio emission. In this {\it radio-based} approach 
$U_{\rm p}$ can be estimated if the source size, distance, radio flux, and radio 
spectral index are known.

\begin{table*}
\caption[] {Star-forming galaxies: the data.}
\begin{flushleft}
\begin{tabular}{ l  l  l  l  l  l  l  l  l  l  l  l }
\hline
\hline
\noalign{\smallskip}

Object & 
$~~D_L^{[1]}$ & 
$~~~r_{\rm s}^{[2]}$ & 
$f_{1\,{\rm GHz}}^{[3]}$ & 
$\alpha_{\rm NT}^{[4]}$ & 
$n_{\rm e,\,th}^{[5]}$ &
$L_{\rm TIR}^{[6]}$ & 
SFR$^{[7]}$ & 
$\nu_{\rm SN}^{[8]}$ & 
$~~~\,M_{\rm gas}^{[9]}$ & 
$~~~\,L_\gamma^{[10]}$ &
~ Notes   \\
               & 
    (Mpc)    & 
   ~(kpc)    & 
   \, (Jy)    &  
               & 
  (cm$^{-3}$)    &
 (erg/s)      & 
 ($M_\odot$/yr) & 
  (yr$^{-1}$)  & 
 $~~\,(M_\odot$) & 
 ~\, (erg/s) &
             \\
\noalign{\smallskip}
\hline
\noalign{\smallskip}

Arp\,220  &\, 74.7 &~ 0.25 &~ 0.3  & 0.65 & 300  & 45.75&  253 & 3.5~~& $9.24^{+0.10}_{-0.11}$ & $<42.25$                & ~~~SB \\
M\,82     &~~3.4   &~ 0.23 &~10.0  & 0.71 & 200  & 44.26&  8.2 & 0.25 & $9.37^{+0.09}_{-0.14}$ & $40.21^{+0.10}_{-0.13}$ & ~~~SB \\
NGC\,253  &~~2.5   &~ 0.20 &~ 5.6  & 0.75 & 400  & 44.23&  7.7 & 0.12 & $9.20^{+0.10}_{-0.11}$ & $39.76^{+0.14}_{-0.19}$ & ~~~SB  \\
Milky Way &~~ --   &~ 4.4  &~  --  & --   & 0.01 & 43.75&  2.5 & 0.02 & $9.81^{+0.12}_{-0.16}$ & $38.91^{+0.12}_{-0.15}$ & quiescent\\ 
M\,31     &\, 0.78 &~ 4.5  &~ 4.0  & 0.88 & 0.01 & 42.98& 0.43 & 0.01 & $9.88^{+0.11}_{-0.15}$ & $38.66^{+0.09}_{-0.10}$ & quiescent\\
M\,33     &\, 0.85 &~ 2.79 &~ 3.30 & 0.95 & 0.03 & 42.68& 0.22 & 0.003& $9.35^{+0.13}_{-0.19}$ & $<38.54$                & quiescent \\
LMC       &\, 0.049&~ 2.4  &~285.0 & 0.84 & 0.01 & 42.45& 0.13 & 0.002& $8.86^{+0.12}_{-0.18}$ & $37.67^{+0.05}_{-0.05}$ & quiescent\\
SMC       &\, 0.061&~ 1.53 &~ 45.3 & 0.85 & 0.01 & 41.45& 0.01 & 0.001& $8.66^{+0.03}_{-0.06}$ & $37.04^{+0.11}_{-0.14}$ & quiescent\\
NGC\,4945 & ~~3.7  &~ 0.22 &~  5.5 & 0.57 & 300  & 44.02&  4.7 &0.1-0.5& $9.64^{+0.10}_{-0.40}$&$40.30^{+0.12}_{-0.16}$  & SB+Sy2  \\
NGC\,1068 & \,16.7 &~ 1.18 &~  6.6 & 0.75 & 300  & 45.05&  50  &0.2-0.4& $9.71^{+0.11}_{-0.19}$& $41.32^{+0.15}_{-0.23}$ & SB+Sy2\\

\noalign{\smallskip}
\hline
\end{tabular}
\smallskip

\noindent
$^{[1]}$Distance (from Ackermann et al. 2012).
\smallskip

\noindent
$^{[2]}$Effective radius of star-forming region. See text. Data are from Persic 
\& Rephaeli 2010 and refs. therein (Arp\,220, M\,82, NGC\,253), Beck \& Gr\"ave 
1982 (M\,31), Tabatabaei et al. 2007 (M\,33), Weinberg \& Nikolaev 2001 (LMC), 
Wilke et al. 2003 (SMC), Moorwood \& Oliva 1994 (NGC\,4945), Spinoglio et al. 2005 
(NGC\,1068).
\smallskip

\noindent
$^{[3]}$1\,GHz flux density. 
Data are from Persic \& Rephaeli 2010 and refs. therein (Arp\,220, M\,82, NGC\,253)), 
Beck \& Gr\"ave 1982 (M\,31), Tabatabaei et al. 2007 (M\,33), Klein et al. 1989 (LMC), 
Haynes et al. 1991 (SMC), Elmouttie et al. 1997 (NGC\,4945), K\"uhr et al. 1981 
(NGC\,1068).
\smallskip

\noindent
$^{[4]}$Non-thermal spectral radio index.
Data are from Persic \& Rephaeli 2010 and refs. therein (Arp\,220, M\,82, NGC\,253), 
Beck \& Gr\"ave 1982 (M\,31), Tabatabaei et al. 2007 (M\,33), Klein et al. 1989 (LMC), 
Haynes et al. 1991 (SMC), Elmouttie et al. 1997 (NGC\,4945), K\"uhr et al. 1981 
(NGC\,1068).
\smallskip

\noindent
$^{[5]}$Thermal electron density. Data are from Roy et al. 2010 (Arp\,220), 
Petuchowski et al. 1994 (M\,82),Kewley et al. 2000 and Corral et al. 1994 (NGC\,253), 
Cox 2005 (Milky Way), Beck 2000 (M\,31), Tabatabaei et al. 2008 (M\,33), Points et al. 
2001 (LMC), Sasaki et al. 2002 (SMC), Spoon et al. 2000 (NGC\,4945), Kewley et al. 
2000 (NGC\,1068).
\smallskip

\noindent
$^{[6]}$Total IR [i.e., $(8-1000)\mu$m] luminosity, in log (from Ackermann et al. 2012).
\smallskip

\noindent
$^{[7]}$Star formation rate, from ${\rm SFR} = L_{\rm IR} / (2.2 \times 10^{43}\, {\rm erg/s} )$ (Kennicutt 1998).
\smallskip

\noindent
$^{[8]}$Core-collapse SN rate. Data are from Persic \& Rephaeli 2010 and references 
therein (Arp\,220, M\,82, NGC\,253), Diehl et al. 2006 (Milky Way), van den Bergh \& 
Tammann 1991 (M\,31, M\,33, SMC, LMC; see also Pavlidou \& Fields 2001), Lenain et al. 
2010 and references therein (NGC\,4945, NGC\,1068). For NGC\,1068 we also computed an 
upper limit to the SN rate ($\nu_{\rm SN} \leq 0.39$) using Mannucci et al.'s (2003) 
formula $\nu_{\rm SN} = (2.4 \pm 0.1) \times 10^{-2} [L_{\rm FIR}/(10^{10} L_\odot)]\,
{\rm yr}^{-1}$, being $f_{\rm FIR}=1.26 \times 10^{-11}(2.58\,f_{60} + f_{100})$ 
erg\,cm$^{-2}$s$^{-1}$ (see Helou et al. 1988) with $f_{60} \simeq 190$\,Jy and 
$f_{100} \simeq 277$\,Jy.
\smallskip

\noindent
$^{[9]}$Gas mass (neutral plus molecular hydrogen: $M_{\rm HI}+M_{\rm H_2}$), in log. 
Data are from: Torres 2004 for Arp\,220; Abdo et al. 2010a for M\,82, NGC\,253, and 
the Milky Way; Abdo et al. 2010b for M\,31 and M\,33; Abdo et al. 2010c for the LMC; 
Abdo et al. 2010d for the SMC; and Lenain et al. 2010 for NGC\,4945 and NGC\,1068.
\smallskip

\noindent
$^{[10]}$High-energy ($>$100\,MeV) $\gamma$-ray luminosity, in log (from Ackermann et 
al. 2012).
\smallskip
%
\end{flushleft}
\end{table*}

In a {\it $\gamma$-based} approach, $U_{\rm p}$ can be obtained from the measured 
GeV-TeV spectral flux, which is mostly due to p-p interactions, as described the 
previous section. Only recently have such measurements become possible, at present 
only for 10 sources.

In the {\it SN method}, with an assumed fraction of SN kinetic energy that is 
channeled into particle acceleration, $U_{\rm p}$ can be estimated if the size of the 
star-forming region and SN rate are known, as well as an estimate for the proton 
residence timescale from the presence (or absence) of a galactic wind emanating 
from the star-forming region. This timescale is largely determined by the advection 
timescale ($\sim 10^5$yr) in SBGs, and by the $\pi^0$-decay timescale ($\sim 10^7$yr) 
in low-SFR (quiescent) galaxies. 

Expanding on our previous work (Persic \& Rephaeli 2010), we show that the three 
methods give consistent results for $U_p$ for a sample of 10 galaxies with widely 
varying levels of star formation activity, from very quiescent to extreme SBs. 
These are the only galaxies of their kind for which $\gamma$-ray data, in addition 
to radio data and SN rates, are available (see Table\,1).

\subsection{Particles and magnetic field}
\label{subsec:1}

The population of NT electrons consists of primary (directly accelerated) and 
secondary (produced via $\pi^\pm$ decays) electrons. While the exact form of the 
steady-state electron energy spectrum is not a single power law, at high energies the 
flattening of the spectrum due to Coulomb losses can be ignored, justifying the 
use of the approximate single power-law form. The combined (primary plus 
secondary) electron spectral density distribution is then 
\begin{eqnarray}
N_{\rm e}(\gamma) ~=~ N_{e,{\rm 0}}\, (1+\chi) ~ \gamma^{-q} , & & 
\label{eq:el_spectrum}
\end{eqnarray}
where the electron Lorentz factor $\gamma$ is in the range $\gamma_1 \leq \gamma 
\leq \gamma_2$, $N_{e,{\rm 0}}$ is a normalization factor of the primary electrons, 
$\chi$ is the secondary-to-primary electron ratio, and $q \geq 2$ is the spectral index.
Ignoring the contribution of low-energy electrons with $\gamma < \gamma_1$, the 
electron energy density is $U_{\rm e} = N_{\rm e,0}\,(1+\chi)\,m_{\rm e} c^2\,
\int_{\gamma_1}^{\gamma_2} \gamma^{1-q} {\rm d}\gamma$, where $\gamma_2$ is an upper 
cutoff whose exact value is irrelevant in the limit of interest, $\gamma_2 >> 
\gamma_1$. For $q>2$ and $\gamma_2 >> \gamma_1$, 
\begin{eqnarray}
U_{\rm e} \simeq N_{e,{\rm 0}}(1+\chi) m_{\rm e}c^2 \gamma_1^{2-q}/(q-2)\,. 
\label{eq:el_en_dens1}
\end{eqnarray}
For a population of electrons (specified by Eq. \ref{eq:el_spectrum}) traversing 
a homogeneous magnetic field of strength $B$ in a region with (a spherically 
equivalent) radius $r_{\rm s}$ located at a distance $d$, standard synchrotron 
relations yield 
\begin{eqnarray}
N_{\rm e,0} (1+\chi) ~=~ 5.72 \times 10^{-15} ~ \psi ~ a(q)^{-1} ~B^{-{q+1 \over 2}} 
~ 250^{q-1 \over 2}
\label{eq:el_spect_norm}
\end{eqnarray}
where the scaled flux is $f_{1\,{\rm GHz}}$ Jy, $a(q)$ is defined and tabulated in, 
e.g., Tucker (1975), and $\psi \equiv ({r_{\rm s} \over 0.1\,{\rm kpc}})^{-3} ({d 
\over {\rm Mpc}})^2 ({f_{1\,{\rm GHz}} \over {\rm Jy}})$. Use of 
Eq.(\ref{eq:el_en_dens1}) then yields
\begin{eqnarray}
U_{\rm e} ~\simeq~ {2.96 \over (1+\chi)} \times 10^{-22} \, 250^{q \over 2}\, \psi \, 
{\gamma_1^{-q+2} \over (q-2) ~a(q)}\, B^{-{q+1 \over 2}}\,.
\label{eq:el_en_dens2}
\end{eqnarray}

In order to compute $U_{\rm e}$ from Eq.(\ref{eq:el_en_dens2}) we need to specify 
$\gamma_1$ and $B$. To do so we make the following assumptions:\\

\noindent
{\it (i)} The low-energy limit of the electron power-law spectrum, $\gamma_1$, marks 
the transition (for decreasing energy) from Coulomb (Rephaeli 1979) to synchrotron 
losses. For an electron of energy $\gamma$, the synchrotron loss rate is 
\begin{eqnarray}
\bigl( {{\rm d}\gamma \over {\rm d}t} \bigr)_{\rm syn} ~=~ - 
1.30 \times 10^{-21} \gamma^2 \biggl( {B \over \mu G} \biggr)^2
{\rm s}^{-1} 
\label{eq:syn_loss}
\end{eqnarray}
whereas the Coulomb loss rate is 
\begin{eqnarray}
\bigl( {{\rm d}\gamma \over {\rm d}t} \bigr)_{\rm coul}  ~=~ - 1.2 \times 10^{-12} 
n_{\rm e,th} ~ \biggl[ 1.0 +{{\rm ln}\,(\gamma/n_{\rm e,th}) \over 75} \biggr]
{\rm s}^{-1}\,. 
\label{eq:coul_loss}
\end{eqnarray}
(Rephaeli 1979). We then simply assume that electrons lose energy via 
Coulomb scattering for $\gamma < \gamma_1$ and via synchrotron cooling for 
$\gamma > \gamma_1$.\\

\noindent
{\it (ii)} The particle energy density is in equipartition with that of the 
magnetic field, $U_{\rm p} + U_{\rm e} = B^2 / 8 \pi$. In terms of the p/e energy 
density ratio, $\kappa$, the equipartition condition is 
$U_{\rm p} [1+ (1+\chi)/\kappa]= B^2/8\pi$, so that 
\begin{eqnarray}
B ~=~ \biggl[ {7.44 \times 10^{-21} \over 1+\chi} \, \bigl[1+{\kappa \over 
1+\chi}\bigr]\, {\gamma_1^{2-q}\, 250^{q/2} \, \psi \over (q-2)~ a(q)} \biggr]^{2 
\over 5+q}  \,. 
\label{eq:equip_B}
\end{eqnarray}
Inserting Eq.(\ref{eq:equip_B}) into Eq.(\ref{eq:syn_loss}) we get $( {{\rm d}\gamma 
\over {\rm d}t})_{\rm syn} \propto \gamma_1^{2(9-q)/(5+q)}$. Once the value of 
$n_{\rm e,th}$ is specified (see Table\,1), by equating Eqs.(\ref{eq:syn_loss},
\ref{eq:coul_loss}) we deduce $\gamma_1$.

The secondary-to-primary electron ratio $\chi$, which appears in Eq.(\ref{eq:equip_B}), 
depends on the injection p/e number ratio, $r_{p/e} = (m_{\rm p} / m_{\rm e})^{(q_{\rm 
inj}-1)/2}$, and on the gas optical thickness to p-p interactions. 
Given the branching ratios in p-p collisions, only a third of these collisions 
produce electrons. The mean free path of CR protons due to p-p interactions 
in a medium of density $n_{\rm p}$ is $\lambda_{\rm pp} = (\sigma_{\rm pp} 
n_{\rm p})^{-1}$; for protons with kinetic energy $T\,\sim$\,{\rm few}\,TeV the 
cross section is $\sigma_{\rm pp} \simeq 50\,{\rm mb} = 5 \times 10^{-26}$\,cm$^2$ 
(Baltrusaitis et al. 1984). For a typical SB ambient gas density $n_{\rm p} \simeq 
150$\,cm$^{-3}$, $\lambda_{\rm pp} \sim 43$\,kpc. The probability for a single 
CR proton to undergo a {\rm pp} interaction in its 3D random walk through 
a region of radius $r_{\rm s} \sim 0.25$\,kpc (also typical of SB nuclei) is then 
$\sqrt{3} \,r_{\rm s} / \lambda_{\rm pp} \simeq 0.01$. Thus, in a typical 
SB environment, characterized by relatively strong non-relativistic shocks 
($q_{\rm inj}=2.2$), the secondary to primary electron ratio is $\chi = \chi_0\, 
\sqrt{3} \,(r_{\rm s}/\lambda_{\rm pp}) \simeq 0.3$. In a more quiescent environment, 
with typical values $n_{\rm p} \simeq 1$\,cm$^{-3}$ and $r_{\rm s} \sim 2.5$\,kpc, 
$\chi \simeq 0.03$. The higher value found in SBs is in approximate 
agreement with results of detailed numerical starburst models for energies 
$\magcir$10\,MeV (plotted in, e.g., Paglione et al. 1996, Torres 2004, De Cea et al. 
2009, Rephaeli et al. 2010).

To compute the p/e energy density, $\kappa$, we assume power-law spectra: {\it (i)} 
The electron spectral index $q_{\rm e}$ is deduced from the measured radio index 
$\alpha$, generally $q_{\rm e} = 2 \alpha + 1$. {\it (ii)} The proton spectral 
index is assumed to be close to the injection value, $q_{\rm p} \sim q_{\rm inj} 
\simeq 2.1-2.2$, for the dense SB environments hosted in the central regions of 
some galaxies, and equal to the leaky-box value, $q_{\rm p} = q_{\rm inj} 
+ \delta \simeq 2.7$ (where $\delta \simeq 0.5$ is the diffusion index) for more 
quietly star forming galaxies. 

Finally, we obtain an explicit expression for $U_{\rm p}$: 
\begin{eqnarray}
U_{\rm p} ~ = ~ {1 \over 8\, \pi} ~ \biggl[ 1+ {1+\chi \over \kappa} \biggr]^{-1} ~ 
\biggl[ {7.44 \times 10^{-21} \over 1+\chi} \, \bigl[1+{\kappa \over 
1+\chi}\bigr]\, {\gamma_1^{2-q}\, 250^{q/2} \, \psi \over (q-2)~ a(q)} \biggr]^{4 \over 5+q} \,. 
\label{eq:U_p}
\end{eqnarray}

Using Eq.(\ref{eq:U_p}), values of $U_{\rm p}$ can be obtained from the 
relevant observational quantities for our sample galaxies; these values are 
listed in Table\,2. The quantities in Eq.(\ref{eq:U_p}) are usually well determined 
for our sample galaxies, except for the p/e energy density ratio $\kappa$, for which 
a spectral index, $q_{\rm p}$, must be assumed. Given its possible values 
(i.e., $q_{\rm p} \simeq 2.1-2.2$ in SB regions, and $q_{\rm p} \simeq 2.1-2.2$ 
for quiescent galaxies), the uncertainty in the spectral index, $\delta q_{\rm p} 
\simeq 0.1$, translates to a factor of $\sim$2 uncertainty on $\kappa$, i.e. 
typically an uncertainty of $\sim$50\% on $U_{\rm p}$ as deduced from 
Eq.(\ref{eq:U_p}).

\subsection{Energetic particles and $\gamma$-ray emission}
\label{subsec:2}

Based on the calculation of $\gamma$-ray emission from SFGs outlined in Section\,1, 
$U_{\rm p}$ can be estimated directly from recent measurements of the nearby galaxies. 
In SBGs, such as M\,82 and NGC\,253, the central SB region (referred to also as the 
source region) with a radius of $\sim 200-300$\,pc is identified as the main site of 
particle acceleration. The injected particle spectrum is assumed to have an index 
$q=2$, the theoretically predicted $N_{\rm p}/N_{\rm e}$ ratio is adopted, and 
equipartition is assumed. A measured radio index $\alpha \simeq 0.7$ in the source 
region implies $q = 2\,\alpha+1 \simeq 2.4$ there, indicating a substantial 
steepening due to diffusion ($D \propto \gamma^{-\delta}$), that cause the 
steady-state particle spectral index to be $q_0+ \delta$ above some break 
energy. The procedure is similar when star formation does not (largely) occur in a 
burst in the nuclear region, but proceeds more more uniformly across the disk. 

For a source with ambient gas number density $n_{\rm gas}$, proton energy density 
$U_{\rm p}$, and volume $V$, the integrated hadronic emission from {\it pp}-induced 
$\pi^0$ decay is
\begin{eqnarray}
L_{\geq E}^{[q]} ~ = ~ \int_V g_{\geq E}^{[q]} ~ n_{\rm gas} ~ U_{\rm p} ~{\rm d}V  
~~ {\rm s}^{-1} \,,
\label{eq:L_hadr}
\end{eqnarray}
with the integral emissivity $g_{\geq \epsilon}^{[\eta]}$ in units of 
photon\,s$^{-1}$(H-atom)$^{-1}$(eV/cm$^3$)$^{-1}$ (Drury et al. 1994). 
Thus, $U_{\rm p}$ can be determined from measurements of $L_{\geq \epsilon}$ 
and $n_{\rm gas}(r)$, and the particles steady-state energy distributions can 
be numerically calculated in the context of the convection-diffusion model. 

In addition to the high-energy detections of the two local SBGs M\,82 and NGC\,253 
(Abdo et al. 2010a, Acciari et al. 2009; Acero et al. 2009), several galaxies with 
low SFR were also detected by the {\it Fermi} telescope. {\it (i)} the Andromeda 
galaxy M\,31 (Abdo et al. 2010b), with $U_{\rm p} \simeq 0.35$\,eV\,cm$^{-3}$; 
{\it (ii)} the Large Magellanic Cloud (LMC) whose average spectrum, either including or excluding the bright star-forming region of 
30\,Doradus, suggests $U_{\rm p} \simeq 0.2-0.3 $\,eV\,cm$^{-3}$ (Abdo et al. 2010c); 
{\it (iii)} SMC for which $U_{\rm p} \simeq 0.15$\,eV\,cm$^{-3}$ was deduced 
(Abdo et al. 2010d). For the Milky Way, the modeling of the Galactic diffuse HE 
emission along the lines outlined above requires an average $U_{\rm p} 
\simeq 1$\,eV\,cm$^{-3}$ (Strong et al. 2010; Ackermann et al. 2011). 
Our values for $U_{\rm p}$ determined from the measured GeV-TeV fluxes are 
listed in Table\,2.

\subsection{Energetic particles and Supernovae}
\label{subsec:3}

The SN origin of energetic particles suggested early on; as a test of this 
hypothesis, we estimate of $U_{\rm p}$ by combining the SN rate with the proton 
residence time, $\tau_{\rm res}$, assuming a fiducial value for the fraction 
of SN kinetic energy that is channeled to particle acceleration. The residence 
time is determined from the p-p interaction time, and the two propagation 
timescales of advection and diffusion:

\noindent
{\it (i)} The energy-loss timescale for {\it pp} interactions, $\tau_{\rm pp} = 
(\sigma_{\rm pp} c n_{\rm p})^{-1}$; for protons with kinetic energy 
$E \magcir 10$\, TeV for which $\sigma_{\rm pp} \simeq 50$\,mb, this 
timescale is 
\begin{eqnarray} 
\tau_{\rm pp} ~\simeq~ 2 \times 10^5 ~ \bigl({ n_{\rm p} \over 100\,{\rm cm}^{-3}}
\bigr)^{-1}~ {\rm yr}\,. 
\label{eq:pp_time}
\end{eqnarray}

\noindent
{\it (ii)} Particle advection out of the disk mid-plane region in a fast SB-driven 
wind occurs on a timescale $\tau_{\rm adv}$ determined from the advection velocity 
for which we adopt (except where noted otherwise) the nominal value $v_{\rm adv} 
\sim 1000$\,km\,s$^{-1}$, deduced from measurements of the terminal outflow 
velocity of $\sim$1600-2200 \,km\,s$^{-1}$ in M\,82 (Strickland \& Heckman 2009; 
see also Chevalier \& Clegg 1985 and Seaquist \& Odegard 1991). For a homogeneous 
distribution of SNe within the SB nucleus of radius $r_{\rm s}$, the advection 
timescale is
\begin{eqnarray}
\tau_{\rm adv} \simeq  
7.5 \times 10^4 ~ \bigl({r_{\rm s} \over 0.3\,{\rm kpc}}\bigr) ~ 
\bigl({ v_{\rm out} \over 1000\,{\rm km\,s}^{-1}}\bigr)^{-1}~ {\rm yr} \,.
\label{eq:adv_time}
\end{eqnarray}

\noindent
{\it (iii)} As noted in the previous section, diffusion is likely to be random 
walk against magnetic field inhomogeneities, with an estimated central diffusion 
coefficient, $D \sim 3 \times 10^{28}$ cm$^2$/s, assuming a magnetic coherence 
scale of $\lambda \sim 1$ pc. Thus, diffusion out of the central 0.5 kpc is 
estimated to occur on a timescale
\begin{eqnarray} 
\tau_{\rm diff} ~\simeq~ 3 \times 10^6 ~ 
\bigl({ r_{\rm s} \over 0.3\,{\rm kpc}} \bigr)^2~
\bigl({ \lambda \over 1\rm pc} \bigr)^{-1}~ {\rm yr} \,.
\label{eq:diff_time}
\end{eqnarray}

Now, since the weighted residence time is 
\begin{eqnarray}
\tau_{\rm res}^{-1} ~=~ 
\tau_{\rm pp}^{-1} + \tau_{\rm adv}^{-1} + \tau_{\rm diff}^{-1} \,,  
\label{eq:res_time}
\end{eqnarray}
it is expected that - under typical conditions in central SB regions - p-p 
collisions and advection, more so than diffusion, effectively determine the 
survival there of energetic protons.

During $\tau_{\rm res}$, the number of SN is $\nu_{\rm SN} \tau_{\rm res}$; 
the kinetic energy deposited by each of these into the ISM is $E_{\rm ej} = 
10^{51}$\,erg (Woosley \& Weaver 1995). Arguments based on the energetic 
particle energy budget in the Galaxy and SN statistics suggest that a 
fraction $\eta \sim 0.05-0.1$ of this energy is available for accelerating 
particles (e.g., Higdon et al. 1998). Thus, the proton energy density can be 
expressed as
\begin{eqnarray}
U_{\rm p}~ = ~ 85 ~ \bigl( {\nu_{\rm SN} \over 0.3\,{\rm yr^{-1}}} \bigr ) ~ 
\bigl( ({\tau_{\rm res} \over 3 \times 10^4\,{\rm yr}} \bigr )  ~ \bigl( {\eta 
\over 0.05}~ {E_{\rm ej} \over 10^{51}\,{\rm erg}} \bigr ) ~ 
\bigl( {r_{\rm s} \over 0.3\,{\rm kpc}} \bigr)^{-3} ~~ {\rm eV~ cm}^{-3}.
\label{eq:CRp_density}
\end{eqnarray}
The resulting values of $U_{\rm p}$ in the sample galaxies are listed in Table\,2.


\begin{table}
\caption[] {Star-forming galaxies: Proton energy densities$^+$.}
\begin{tabular}{ l  l  l  l  l  l  l }
\hline
\hline

\noalign{\smallskip}
Object & $\gamma$-ray & radio & SN & other & $~~r_{\rm s}$ & $\tau_{\rm res}$ \\
       &   meth.      & meth. &meth.& meth.&   (kpc)       & (yr) \\
\noalign{\smallskip}
\hline
\noalign{\smallskip}
Arp\,220              & ~\,   --        & 1027 & 1142 & ~\,  --           & ~0.25 &  2.0E+4 \\
M\,82                 & $250^{\rm a,c}$ & ~250 & ~234 & ~\,  --           & ~0.23 &  4.5E+4 \\
NGC\,253              & $220^{\rm b,c}$ & ~230 & ~213 & ~\,  --           & ~0.20 &  6.7E+4 \\
Milky Way             & $~~1^{\rm d}  $ &~\,-- & ~1.2 & ~\, $1^{\rm j}$   & ~4.4  &  2.0E+7 \\
                      & $~~6^{\rm e}  $ &~\,-- & ~\,5 & ~\,  --           & ~0.2  &  2.5E+6 \\
M\,31                 & $0.36^{\rm f} $ &~0.22 & 0.7  & ~\,  --           & ~4.5  &  2.5E+7 \\ 
M\,33                 & $<0.43^{\rm f}$ &~0.38 & 0.7  & ~\,  --           & ~2.8  &  2.0E+7 \\
LMC              & $0.21-0.31^{\rm g} $ &~0.22 & 0.4  & ~\,  --           & ~2.5  &  4.4E+7 \\
SMC                   & $0.15^{\rm h} $ &~0.39 & 1.1  & ~\,  --           & ~1.5  &  1.4E+7 \\
NGC\,4945             & $200^{\rm i}  $ & ~201 & 215  & ~\,  --           & ~0.22 &  4.5E+4 \\
NGC\,1068             & ~\,   --        & ~~65 & ~61  & ~\,  --           & ~1.2  &  1.0E+6 \\

\noalign{\smallskip}
\hline
\end{tabular}
\smallskip

$^+$ Values are in eV\,cm$^{-3}$. 
\smallskip

\noindent
{\it (a)} Acciari et al. (2009; see also Persic et al. 2008, De Cea et al. 2009). 
{\it (b)} Acero et al. (2009).
{\it (c)} Abdo et al. (2010a).
{\it (d)} Strong et al. (2010).
{\it (e)} Aharonian et al. (2006).
{\it (f)} Abdo et al. (2010b), and Drury et al. (1994) for M\,33.
{\it (g)} Abdo et al. (2010c).
{\it (h)} Abdo et al. (2010d).
{\it (i)} Lenain et al. (2010).
{\it (j)} Webber (1987).
\end{table}

\section{Discussion}
\label{sec:4}

Detections of VHE $\gamma$-ray emission associated with ongoing star formation in 
M\,82 and NGC\,253 add significant new insight on the enhanced energetic electron 
and proton contents in SBGs, and on their propagation in disks of spiral galaxies. 
The common star-driven nature of NT phenomena in the wide class of non-AGN SFGs 
implies that energetic particle densities and radiation fields are self-similarly 
scaled with SF activity, from quiescent systems to intense SBGs, in spite of the 
wide range of intrinsic physical conditions in these systems.

We have briefly outlined our treatment of the steady state spectro-spatial 
distributions of energetic electrons and protons in SFGs, exemplified in the case of 
the two nearby SBGs M\,82 and NGC\,253. This approach is based on a numerical solution 
of the diffusion-convection equation for particle distribution functions, following 
the evolution from the acceleration sites throughout the disk as the particles lose 
energy and propagate outward. Key observational normalization is based on measurements 
of radio synchrotron emission which, through an initial (theoretically assumed) 
$N_{p}/N_{e}$ ratio provides also the normalization of the proton component. Assuming 
equipartition then allows to relate the local value of the mean magnetic field to 
the particle energy densities. The numerical solution of the diffusion-convection 
equation is based on an iterative procedure to determine the particle densities and 
mean field strength in the galactic center, and evolve these quantities by accounting 
for all relevant energy losses, and normalizing central values of these quantities 
by fitting to the measured radio spectrum from the central galactic (or SB) region. 
The quantitative viability of this approach is confirmed by the good agreement 
between the predicted high energy emission from M\,82 and NGC\,253 and measuremnts 
with {\it Fermi}, H.E.S.S., and VERITAS.

Significant detections of the NT emission from above two SBGs at lower (below 100 keV) 
X-ray energies, as would be expected by the currently operational NuSTAR telescope, 
will provide additional spectral coverage that will allow separating out the 
spectral electron and proton components. NuSTAR is the first X-ray telescope with 
capability to resolve this emission; if this is indeed achieved, important new 
insight will be gained on the evolution of the electron spectro-spatial distribution 
across the disks of these nearby SBGs.

The three methods we have discussed to estimate energetic particle energy densities 
are clearly not independent. The $\gamma$-ray method and the radio method are coupled 
through the p/e ratio at injection, through the secondary-to-primary electron ratio, 
and through the imposed condition of particle-field equipartition. The SN method is 
not independent of the $\gamma$-ray method either, because both depend on the proton 
residence time, although - unlike the $\gamma$-ray and radio methods - it does not 
depend on the particle radiative yields but on the statistics of core-collapse SN. 
Also, the three methods do not stand on equal footing: with the $\gamma$-ray, radio, 
and SN methods we, respectively, either {\it measure}, {\it infer}, and or 
{\it estimate} the value of $U_{\rm p}$. A substantial agreement among estimates 
based on the three methods is found for most of the galaxies in Table\,1. The 
only exceptions are the SMC and NGC\,1068. As for the former, the proton 
confinement volume could be small, so that most particles diffuse out to 
intergalactic space (Abdo et al. 2010d). If so, the $\gamma$-ray method 
yields the (lower) {\it actual} proton energy density, whereas the radio 
and SN methods estimate the (higher) {\it produced} amount. NGC\,1068 hosts a 
prototypical Seyfert-2 nucleus (e.g., Wilson \& Ulvestad 1982) surrounded by a 
spherical circumnuclear SB shell with external radius of 1.5\,kpc and thickness 
0.3\,kpc, and mass $3.4 \times 10^9 M_\odot$ (Spinoglio et al. 2005); its implied 
energy density is $U_{\rm p} \approx 65$\,eV\,cm$^{-3}$ from both the 
radio and SN methods. [We note that Lenain et al. (2010) suggested that the HE 
emissions of the above two SB and Sey\,II galaxies NGC\,4945 and NGC\,1068 are 
powered by, respectively, star formation and AGN activity.]

A debated aspect of proton energy loss and propagation times is whether the former 
is shorter than the latter; if so, the system is said to be a `proton calorimeter' 
(e.g., Lacki et al. 2010, 2011). No galaxy in the above sample is found to be in 
the calorimetric limit; the two SBGs M\,82 and NGC\,253 would seem to be only 
marginally close to this limit. The presence of fast, SB-driven galactic 
winds advecting energetic particles out of the disk seems to be a ubiquitous 
feature in SBGs, limiting the degree at which they can be calorimetric. More 
generally, it is known that energetic particles do diffuse out of non-AGN SFGs, 
as evidenced also by significant radio emission (Ferrari et al. 2008), and possibly 
also high energy NT X-ray emission (Rephaeli et al. 2008) from large central 
regions of galaxy clusters. The estimated intracluster particle (and indeed 
also the magnetic) energy densities are sufficiently high, suggesting origin in 
the cluster galaxies.

Based on the reasonable hypothesis that local SB galaxies resemble young galaxies  
which were particularly abundant in the early universe, their contributions to 
the X-$\gamma$-ray backgrounds are of obvious interest (e.g., Rephaeli et al. 1991). 
Calculations of the superposed emission from SBGs (Pavlidou \& Fields 2002, 
Persic \& Rephaeli 2003, Thompson et al. 2007, Fields et al. 2010, Lacki et al. 
2011, Steckers \& Venters 2011) indicate that this emission constitutes at least 
a tenth of these backgrounds.


\end{document}